\documentclass[pra,twocolumn,showpacs,floatfix,amsfonts]{revtex4}
\usepackage{bm}
\usepackage{graphicx}  
\usepackage{amsmath}     
\usepackage{amsfonts}  
\usepackage{amssymb}  
\usepackage{dcolumn}

\newcommand{\bra}[1]{\langle #1 | \,}
\newcommand{\ket}[1]{\, | #1 \rangle}

\newcommand{\be}{\begin{equation}}
\newcommand{\ee}{\end{equation}}
\newcommand{\bea}{\begin{eqnarray}}
\newcommand{\eea}{\end{eqnarray}}

\def\unity{\openone}

\newcommand{\Mr}{\mathfrak{M}}
\newcommand{\Er}{\mathfrak{E}}

\newcommand{\PosInt}{\mathbb{N}}
\newcommand{\Int}{\mathbb{Z}}
\newcommand{\Rat}{\mathbb{Q}}

\newcommand{\Set}{\mathbb{S}}
\newcommand{\Hil}{\mathbb{H}}

\newcommand{\Dr}{\mathbb{D}}

\begin{document}
\title{Applications of single-qubit rotations in quantum public-key cryptography}
\author{Georgios M. Nikolopoulos}
\affiliation{Institute of Electronic Structure and Laser, FORTH, 
P. O. Box 1527, Heraklion 711 10, Crete, Greece}

\date{\today}

\begin{abstract}
We discuss cryptographic applications of single-qubit rotations from the perspective 
of trapdoor one-way functions and public-key encryption. 
In particular, we present an asymmetric cryptosystem whose security relies 
on fundamental principles of quantum physics. 
A quantum public key is used for the encryption of messages while decryption 
is possible by means of a classical private key only. 
The trapdoor one-way function underlying the 
proposed cryptosystem maps integer numbers to quantum states of a qubit 
and its inversion can be infeasible by virtue of the Holevo's theorem. 
\end{abstract}


\maketitle

\section{Introduction} 
Modern public-key (or else asymmetric) cryptography relies on numerical 
trapdoor one-way functions, i.e., functions that are ``easy'' to compute, but 
``hard'' to invert without some additional information (the so-called 
trapdoor information) \cite{book2}. The main characteristic 
of these mathematical objects is that they provide the legitimate 
users with a tractable problem, while at the same time any 
unauthorized user (adversary) has to face a computationally 
infeasible problem. This barrier between legitimate users and adversaries, 
due to complexity of effort, is the key idea 
behind  most of the known public-key cryptosystems. 
Each participant in such a cryptosystem has to have a personal key consisting 
of two parts, i.e., the public and the secret (also known as private) part. 
Messages are encrypted with use of the public key and the decryption of the 
resulting ciphertext is possible by means of the private key. 

The security of conventional public-key cryptography relies on the 
hardness of some computational problems 
(e.g., integer factorization problem, discrete logarithm problem, etc). 
These numerical problems are considered to be 
good candidates for one-way functions (OWFs), 
and this belief relies mainly on the large 
amount of resources (computing power and time) required for their 
solution using the best known algorithms.  
Nevertheless, the fact that the existence of numerical OWFs 
has not been proved rigorously up to now, makes all of the known public-key 
cryptosystems vulnerable to any future advances in algorithms and 
hardware (e.g., the construction of a quantum computer). 

In contrast to the computational security offered by conventional 
public-key schemes, there exist symmetric cryptosystems (e.g., one-time 
pad) which offer provable security provided that a {\em secret truly 
random} key is shared between the entities who wish to 
communicate. 
Today, the establishment of such a key between two parties 
can be achieved by means of quantum key-distribution (QKD) protocols \cite{RMP-QKD}. 
By virtue of fundamental principles of quantum mechanics that do not 
allow passive monitoring and cloning of unknown quantum states \cite{book3},  
QKD protocols provide a solution to the {\em key-distribution problem} 
even in the presence of the most powerful adversaries.
Nevertheless, the key management remains one of the main drawbacks of 
symmetric encryption schemes \cite{book2}. In particular, the problem pertains to large 
networks where each entity needs a secret key with every other entity.  
Hence, the total number of secret keys scales quadratically with the number of 
users in the network. 

One solution to the {\em key-management problem} is the use of 
an {\em unconditionally trusted} third party which is burdened 
with the key management and acts as a key-distribution center (KDC). 
The main problem with this solution, however, is that the KDC itself becomes an 
attractive target, while a compromised KDC renters immediately 
all communications insecure. An alternative solution to the key-management problem 
is provided by conventional public-key cryptosystems which are very flexible but, 
as we discussed earlier, offer computationally security only. 

Clearly, an ideal solution to both of the key-distribution and management problems 
is a quantum public-key (asymmetric) cryptosystem, which combines the provable 
security of QKD protocols with the flexibility of conventional public-key 
encryption schemes.  
The development of such a cryptosystem, however, requires the 
existence of quantum trapdoor OWFs.  
In particular, the one-way property of these functions has to rely on 
fundamental principles of quantum theory, rather than unproven 
computational assumptions.  

To the best of our knowledge, the number of related theoretical 
investigations is rather small, and all of them pertain to 
a futuristic scenario where all of the parties involved 
(legitimate users and adversaries) possess quantum computers.    
The concept of quantum OWF was first introduced in \cite{Buh01,Got01}, 
where the authors demonstrated that such a function can be obtained by mapping 
classical bit-strings to quantum states of a collection of qubits. 
Nevertheless, these two papers do not pertain directly to public-key encryption,  
but rather to quantum fingerprinting \cite{Buh01}, and digital signatures \cite{Got01,And06}.
Later on, Kawachi {\em et al.} \cite{Kaw05} investigated the cryptographic properties 
of the distinguishability problem between two random coset states with 
hidden permutation. This problem can be viewed as a quantum extension 
of the distinguishability problems between two probability distributions 
used in conventional cryptography \cite{book2}. 
Finally, besides quantum OWFs there have 
been also investigations on OWFs which rely on ``hard'' problems 
appearing in other areas of physics such as statistical physics \cite{Kab00}, 
optics \cite{Pen00}, and mesoscopic physics of disordered media \cite{Pap02}.   

In this paper we establish a theoretical framework for quantum public-key 
encryption based on qubit rotations. In particular, we explore the 
trapdoor and one-way properties of functions that map integer numbers 
onto single-qubit states. Moreover, we present an asymmetric cryptosystem 
which is provably secure even against powerful quantum eavesdropping strategies. 

\section{Quantum trapdoor (one-way) functions}
\label{sec2}
In this section we introduce the notion of the quantum trapdoor OWF, 
that maps integer numbers to quantum states of a physical system. 
The discussion involves a scenario where all of the parties
(legitimate users and adversaries) possess quantum computers and are only 
limited by the laws of physics. 

\subsection{Definition and properties}
{\em Definition.} Consider two sets $\Set$ and $\Rat$ which involve 
numbers and quantum states of a physical system, respectively. 
A quantum OWF is a map $\Mr: \Set\mapsto \Rat$, 
which is ``easy'' to perform, but ``hard'' to invert.  
A quantum OWF whose inversion becomes feasible by means of some 
information (trapdoor information) is a quantum trapdoor OWF. 

Throughout this work we will focus on quantum trapdoor OWFs  
whose input is an integer $s\in \Int_n:=\{0,1,\ldots,n-1|n\in\PosInt\}$, 
and its output is the 
state of a quantum system, say $\ket{\phi_s}$. To elaborate further on the 
terms ``easy'' and ``hard'', consider a quantum system initially prepared in some state 
$\ket{0}$ and let $\Hil$ be the corresponding Hilbert space. 
For a randomly chosen $s\in \Int_n$ we apply an operation 
$\hat{{\cal O}}(s):\Hil\mapsto \Hil $ on the system, which changes the initial state 
$\ket{0}\to\ket{\phi_s}=\hat{{\cal O}}(s)\ket{0}$. 
The set of all possible output states of the quantum OWF is 
$\Rat\equiv \{\ket{\phi_s}|s\in \Int_n\}$, and belongs to $\Hil$. 
If the map $\Mr: \Int_n\mapsto \Rat$ is a {\em bijection} there is a unique 
$s\in\Int_n$ such that $\ket{0}\to\ket{\phi_s}$, 
i.e., $\Mr$ is one-to-one and $|\Int_n|=|\Rat|$.

The map $s\mapsto\ket{\phi_s}$ must be ``easy'' to compute in the sense that, for a given 
$s\in\Int_n$,  the transformation on the system's state $\ket{0}\to\ket{\phi_s}$, 
can be performed efficiently on a quantum 
computer with polynomial resources. 
On the other hand, in order for the map $s\mapsto\ket{\phi_s}$ to serve as a quantum OWF,  
its inversion must be a ``hard'' problem by virtue of fundamental principles of 
quantum mechanics. In other words, given a state $\ket{\phi_s}$ chosen at random from $\Rat$, 
there is no efficient quantum algorithm that succeeds in performing the inverse map 
$\ket{\phi_s}\mapsto s$ (i.e., recovering the integer $s$ from the given 
state $\ket{\phi_s}$) with a non-negligible probability.
 
Actually, by definition the inversion of a quantum OWF 
is a hard problem for everyone (legitimate users and eavesdroppers). 
For cryptographic applications, however, authorized users 
should be able to identify the state of the quantum system, and thus inverse the 
map  $s\mapsto\ket{\phi_s}$, more efficiently than any unauthorized party. 
Hence, it is essential to introduce a trapdoor information which makes the 
inversion of the map computationally feasible for anyone who possesses it.

Having introduced the notion of quantum trapdoor OWFs in a rather general 
theoretical framework, in the following we  specialize the present 
discussion to a particular family of such functions based on single-qubit 
rotations. 

\subsection{A quantum trapdoor function based on single-qubit rotations}
For the sake of simplicity, we will present our quantum trapdoor OWF in the 
context of single-qubit states lying on the $x-z$ plane of the 
Bloch-sphere. The main idea can be easily extended to qubit states 
that lie on the three-dimensional Bloch sphere. 

Let us denote by $\{\ket{0_z},\ket{1_z}\}$ the eigenstates 
of the Pauli operator $\hat{{\cal Z}}=(\ket{0_z}\bra{0_z}-\ket{1_z}\bra{1_z})$, which 
form an orthonormal basis in the Hilbert 
space of a qubit $\Hil_2$. A general qubit state lying on the $x-z$ plane 
can be written as  $\ket{\psi(\theta)}=\cos\left (\theta/2\right )\ket{0_z}
+\sin\left (\theta/2\right )\ket{1_z}$, 
where $0\leq \theta< 2\pi$. 
Hence unlike the classical bit which can store a discrete 
variable taking only two real values (that is ``0'' and ``1''), 
a qubit may represent a continuum of states on the $x-z$ Bloch 
plane. Introducing the rotation operator about the $y$ axis, 
$\hat{\cal R}(\theta)=e^{-{\rm i}\theta \hat{\cal  Y}/2}$ with 
$\hat{\cal Y}={\rm i}(\ket{1_z}\bra{0_z}-\ket{0_z}\bra{1_z})$, we may 
alternatively write $\ket{\psi(\theta)}=\hat{\cal R}(\theta)\ket{0_z}$.

The input of the proposed quantum trapdoor function is a random integer 
$s$ uniformly distributed over $\Int_{2^n}$ with $n\in\PosInt$, and a qubit initially 
prepared in $\ket{0_z}$. Thus, $n$-bit strings suffice as labels to 
identify the input $s$ for fixed $n$.  
For given values of $n\in\PosInt$ and $s\in\Int_{2^{n}}$, the qubit state 
is rotated by $s\theta_n$ around the $y$-axis with $\theta_n=\pi/2^{n-1}$.
Hence, for some fixed $n\in\PosInt$, the output of the OWF  
pertains to the class of states 
$\Rat_{n}=\{\ket{\psi_s(\theta_n)}| s\in\Int_{2^n}, \theta_n=\pi/2^{n-1}\}$,
with 
\bea
\ket{\psi_s(\theta_n)}&\equiv&\hat{\cal R}(s\theta_n)\ket{0_z}\nonumber\\
&=&\cos\left (\frac{s\theta_n}{2}\right )\ket{0_z}
+\sin\left (\frac{s\theta_n}{2}\right )\ket{1_z}.
\label{1s}
\eea
Clearly, both of the input and output sets (i.e., $\Int_{2^n}$ and 
$\Rat_n$, respectively) remain unknown if $n$ is not revealed. 

For a given pair of integers $\{n,s\}$, the function $s\mapsto\ket{\psi_s(\theta_n)}$ 
is easy to compute since it involves single-qubit rotations only. In particular, 
it is known that any single-qubit operation can be simulated to an arbitrary accuracy 
$\epsilon>0$, by a quantum algorithm involving a universal set of gates 
(i.e., Hadamard, phase, controlled-NOT, and $\pi/8$ gates) \cite{book3}. 
Moreover, this simulation is efficient since its implementation 
requires an overhead of resources that scales polynomially with $\log(\epsilon^{-1})$. 
 
Inversion of the map $s\mapsto\ket{\psi_s(\theta_n)}$ means to recover $s$ 
from a given qubit state $\ket{\psi_s(\theta_n)}$ chosen at random from an  
{\em unknown} set $\Rat_n$. Nevertheless, let us consider for the time being that 
$n$ is known. In this case, the inversion of the map reduces to the problem 
of discrimination between various non-orthogonal states chosen at random 
from a known set $\Rat_n$. 
The number of non-orthogonal states increases as we increase $n$, 
whereas for $n>>1$ we have for the nearest-neighbor overlap 
$\bra{\psi_s(\theta_n)}\psi_{s+1}(\theta_n)\rangle=\cos(\theta_n/2)\to 1$.
Hence, a projective von Neumann measurement cannot distinguish between all of the 
states for $n>>1$, since the number of possible outcomes in such a measurement 
is restricted by the dimensions of the state space of the system 
(i.e., qubit in our case). 

One has therefore, to resort to more general measurements which can be always 
described formally by a positive operator-valued measure (POVM) involving a number 
of non-negative operators \cite{book3}. In the theoretical framework of POVMs, 
an input state is associated with a particular outcome of the measurement, 
while optimization is typically performed with respect to 
various quantities (e.g., probability of inconclusive results, 
mutual information, conditional probabilities, etc). 
It is worth noting, however, 
that some of these strategies are not applicable for the states of the set 
$\Rat_n$, since they are not linearly independent when $n>2$ 
(e.g., see Ref. \cite{ChefPLA98}).  
In any case, according to Holevo's theorem \cite{book3}, 
the classical information that can be 
extracted from a single qubit by means of a POVM is at most $1$ bit, 
whereas $n$ bits required to identify the randomly chosen $s\in\Int_{2^n}$ 
for fixed $n$. Hence, we see that for a given $n\gg 1$ the map 
$s\mapsto\ket{\psi_s(\theta_n)}$ acts as a quantum OWF that is ``easy'' 
to perform but hard to invert. Actually, the inversion may become even  
harder if $n$ is not publicly announced, thus rendering the sets from which $s$ and 
$\ket{\psi_s(\theta_n)}$ are chosen (that is, $\Int_{2^n}$ and $\Rat_n$, respectively) 
practically unknown (see also discussion in Sec. \ref{secIV}).

The map $s\mapsto\ket{\psi_s(\theta_n)}$ may also act as a trapdoor OWF 
when it involves two consecutive rotations. 
To demonstrate this fact, let us assume that after $\hat{\cal R}(s\theta_n)$, 
a  second rotation 
$\hat{\cal R}(m\theta_n)$ is applied to the same qubit, 
with a randomly chosen integer $m\in\Int_{2^n}$ such that $s+m=c\mod 2^{n}$. 
The state of the qubit after the second rotation becomes  
$\ket{\psi_c(\theta_n)}=\hat{\cal R}(c\theta_n)\ket{0_z}=
\hat{\cal R}(m\theta_n)\hat{\cal R}(s\theta_n)\ket{0_z}$. 
Having access to the qubit before and after the second rotation 
(i.e., given the qubit states $\ket{\psi_s(\theta_n)}$ and $\ket{\psi_{c}(\theta_n)}$),  
we are interested in deducing $m$. This task, however, requires 
substantial information on both of the numbers $s$ and $c$, which is not possible 
for $n\gg 1$. More precisely, as discussed earlier, in this case only negligible 
information can be extracted from the state $\ket{\psi_s(\theta_n)}$ about the randomly 
chosen $s$, which thus remains practically unknown. 
Hence, irrespective of the amount of information one may have 
on $c$, the number $m$ will also remain unknown. 
The one-way and trapdoor properties of the map $s\mapsto\ket{\psi_s(\theta_n)}$ 
will become clearer in the following, through the security analysis 
of an asymmetric quantum encryption scheme.

\section{Quantum public-key encryption}
\label{secIII}
In this section we introduce an asymmetric cryptosystem based on the quantum 
trapdoor OWF presented in Sec. \ref{sec2}. 
In analogy to classical asymmetric cryptosystems, in the proposed protocol 
the encryption and the decryption keys are different.  
In the following we describe the three stages of the protocol. 

\label{secIIIa}
{\em Stage 1 --- Key generation.} Each user participating in the cryptosystem 
generates a key consisting of a private part $d$, and a public part $e$, 
as determined by the following steps. 
\begin{enumerate}
\item Choose a random positive integer $n\gg 1$.

\item Choose a random integer string ${\bf s}$ of length $N$ i.e.,  
${\bf s}=(s_1,s_2,\ldots,s_N)$, with $s_j$ chosen independently 
from $\Int_{2^n}$.
 
\item Prepare $N$ qubits in the state $\ket{0_z}^{\otimes N}$. 

\item Apply a rotation $\hat{\cal R}^{(j)}(s_j\theta_n)$ on the $j$th qubit,  
with $\theta_n=\pi/2^{n-1}$. Thus, the state of the $j$th qubit becomes 
$\ket{\psi_{s_j}(\theta_n)}_j=\hat{\cal R}^{(j)}(s_j\theta_n)\ket{0_z}$, 
and is of the form (\ref{1s}). 

\item The private key is $d=\{n,{\bf s}\}$, while the public key is 
$e=\{N,\ket{\Psi_{{\bf s}}^{\rm (pk)}(\theta_n)}\}$, with the $N$-qubit state 
$\ket{\Psi_{{\bf s}}^{\rm (pk)}(\theta_n)}\equiv\bigotimes_{j=1}^{N}
\ket{\psi_{s_j}(\theta_n)}_j$.

\end{enumerate}
Clearly in the proposed protocol, the private key is classical whereas 
the public key is quantum as it involves the state of  $N$ qubits. 
Moreover, note that each user may produce multiple copies of his/her own public 
key as the quantum state is known, and thus its copying does not violate the 
no-cloning theorem. 

{\em Stage 2 --- Encryption.} Assume now that one of the users (Bob) 
wants to send Alice an $r$-bit message 
${\bf m}=(m_1,m_2,\ldots,m_r)$, with $m_j\in\{0,1\}$ and $r\leq N$. 
To encrypt the message, he should do the following steps without altering 
the order of the public-key qubits:
\begin{enumerate}
\item Obtain Alice's {\em authentic} public key $e$. 
If $r>N$, he should ask Alice to increase the length of her public key.

\item Encrypt the $j$th bit of his message, say $m_j$, by applying the 
rotation $\hat{\cal R}^{(j)}(m_j\pi)$ on the corresponding 
qubit of the public key, whose state becomes 
$\ket{\psi_{s_j,m_j}(\theta_n)}_j=
\hat{\cal R}^{(j)}(m_j\pi)\ket{\psi_{s_j}(\theta_n)}_j$. \\

\item The quantum ciphertext (or else cipher state) is the new state 
of the $N$ qubits, i.e., $\ket{\Psi_{{\bf s},{\bf m}}^{\rm (c)}(\theta_n)}=
\bigotimes_{j=1}^N\ket{\psi_{s_j,m_j}(\theta_n)}_j$,
and is sent back to Alice. 

\end{enumerate}

Note that, at the end of the encryption stage, the message has been encoded in the 
first $r$ qubits of the cipher state. Thus, in the decryption stage 
Alice may focus on this part of the cipher state, discarding the remaining 
$N-r$ qubits, which do not carry any additional information.

{\em Stage 3 --- Decryption.} 
To recover the plaintext ${\bf m}$ from the cipher state
$\ket{\Psi_{{\bf s},{\bf m}}^{\rm (c)}(\theta_n)}$, Alice has to perform
the following steps. 
\begin{enumerate}
\item Undo her initial rotations, i.e., to apply 
$\hat{R}^{(j)}(s_j\theta_n)^{-1}$ 
on the $j$-th qubit of the ciphertext.

\item Measure each  qubit of the ciphertext in the basis $\{\ket{0_z},\ket{1_z}\}$. 
\end{enumerate}

In discussing the decryption stage, we would like to point out that the above two steps 
are basically equivalent to a von Neumann measurement which projects the $j$th qubit 
onto the basis 
$\{\ket{\psi_{s_j}(\theta_n)}, \hat{\cal R}(\pi)\ket{\psi_{s_j}(\theta_n)}\}$. 
Moreover, it is worth recalling here that 
$\hat{R}^{(j)}(\alpha)^{-1}=\hat{R}^{(j)}(\alpha)^{\dag}=
\hat{R}^{(j)}(-\alpha)$, while different rotations around the same axis 
commute, i.e., $[\hat{\cal R}^{(j)}(\alpha),\hat{\cal R}^{(j)}(\beta)]=0$.

\section{Security}
\label{secIV}
The primary objective of an adversary (eavesdropper) is to recover the plaintext 
from the cipher state intended for Alice. On the other hand, there is always 
a more ambitious objective pertaining to the recover of the private key from 
Alice's public key. A cryptosystem is considered to be broken with accomplishment 
of any of the two objectives, but in the latter case the adversary has 
access to all of the messages sent to Alice. In this section we discuss various 
security issues related to the encryption scheme of Sec. \ref{secIII}

\subsection{Distribution of public keys}
In contrast to symmetric cryptosystems, in an asymmetric cryptosystem 
a KDC is burdened with the distribution of public keys whose secrecy is not 
required. Nevertheless, the KDC has to verify still the public key of each 
entity participating in the cryptosystem. Typically, in conventional 
cryptography the outcome of this verification is a public certificate 
which consists of two parts; a data part which contains 
the public key as well as information about its owner, and the 
verification part with the signature of the KDC over 
the data part. Hence, such a certificate essentially guarantees the 
authenticity, or else integrity, of the public key of each entity. 

Authentication is a crucial requirement for secure, classical or quantum,  
encryption schemes since without it any encryption scheme is vulnerable to 
an impersonation attack \cite{book2}. 
In modern cryptography, secrecy (confidentiality) and 
authenticity are treated as distinct and independent cryptographic goals \cite{book2}. 
In particular, public-key encryption aims at confidentiality 
whereas other cryptographic goals (such as data integrity, authentication, and 
non-repudiation)  are provided by other cryptographic primitives including message 
authentication codes, digital signatures, and fingerprints. 
Following the same attitude, throughout this section we focus on the security provided  
by the quantum encryption scheme under consideration. 

To emphasize, however, the importance of authenticity, in the encryption 
stage of the protocol described in Sec. \ref{secIII} it is explicitly stated that Bob 
should be able to obtain an 
authentic copy of Alice's public key. A quantum digital signature scheme 
for authentication purposes was proposed in \cite{Got01}, and relies on mapping 
classical bit-strings to multi-qubit states.  
We believe that the main results of \cite{Got01} can be also adapted to the 
single-qubit OWF discussed here. 
Nevertheless, the creation of public certificates for quantum keys is 
not an easy task, since digitally signing an unknown qubit state is not possible 
\cite{Cur02}.  
In any case, authentication of quantum messages remains an interesting question 
in the field of quantum cryptography, but it is beyond the scope of this paper.

\subsection{Secrecy of the private key}
\label{secIVB}
The private key of each entity consists of two parts i.e., $d=\{n,{\bf s}\}$. 
The first part is a randomly chosen positive integer with the only constraint being 
$n\gg 1$. Nevertheless, to present quantitative estimates 
on the entropy of the private key, in the following we consider that 
$n$ is uniformly distributed over a finite interval 
$\tilde{\PosInt}=[n_{\rm l},n_{\rm u}]$, with $n_{\rm l}\gg 1$. Thus, the entropy of the 
first part of the private key is $H(n)=\log_2(|\tilde{\PosInt}|)$, 
where $|\tilde{\PosInt}|$ denotes the number of elements in $\tilde{\PosInt}$.
The second part of the private key involves a random integer string 
${\bf s}$, which is encoded on the state of the $N$ qubits of the 
public key.  For a given value of $n$, say $n=\nu$, each random 
element of ${\bf s}$ is chosen independently and has a uniform distribution 
over $\Int_{2^\nu}$. Hence the sting ${\bf s}$ is also uniformly 
distributed over 
$\Int_{2^\nu}^N\equiv\{(a_1,a_2,\ldots,a_N)|a_j\in\Int_{2^\nu}\}$, 
and its entropy is given by $H({\bf s}|n=\nu)=N \nu$. 
The entropy of the entire private key is given by  
the joint entropy $H(n,{\bf s})$, i.e.,     
$H(d)= H(n)+H({\bf s}|n)=\log_2(|\tilde{\PosInt}|)+\sum_{\nu\in\tilde{\PosInt}} p(\nu)H(s|n=\nu)
=\log(|\tilde{\PosInt}|)+N(n_{\rm u}+n_{\rm l})/2$. 

Let us estimate now the classical information one may extract from the quantum 
public key. 
For a given value of $n=\nu$, the $j$th element of ${\bf s}$ is chosen at random 
from $\Int_{2^\nu}$, and the corresponding qubit of the public key is prepared in the 
pure state $\ket{\psi_{s_j}(\theta_\nu)}_j$. 
From an adversary's point of view, however, who does not have access to  
$s_j$, the $j$th qubit of the public key is prepared in a pure state chosen 
at random from the set 
$\Rat_{n=\nu}=\{\ket{\psi_{s_j}(\theta_\nu)}~|~s_j\in\Int_{2^\nu};\theta_\nu=\pi/2^{\nu-1}\}$, 
with all the states being equally probable. Accordingly, one can easily show that for 
$\nu>2$, the density operator for the $j$th qubit is of the form 
\begin{subequations}
\be
\sigma_{\rm pk}^{(j)}(\theta_{n=\nu})=\frac1{2^{\nu}}\sum_{s_j=0}^{2^{\nu}-1}
\ket{\psi_{s_j}(\theta_\nu)}_{jj}\bra{\psi_{s_j}(\theta_\nu)}=\frac{\unity}{2}.
\label{mix}
\ee
Summing over all possible values of $n$ and 
taking into account its uniform distribution over $\tilde{\PosInt}$, we obtain 
$\rho_{\rm pk}^{(j)}=|\tilde{\PosInt}|^{-1}\sum_{n} \sigma_{\rm pk}^{(j)}(\theta_n)=\unity/2$.
Moreover, each qubit is prepared independently of the others, and thus the 
state of the entire public key reads 
\be
\rho_{\rm pk}=\sum_{n\in{\tilde\PosInt}} \sum_{{\bf s}\in\Int_{2^n}} p(n,{\bf s})
\ket{\Psi_{{\bf s}}^{\rm (pk)}(\theta_n)}\bra{\Psi_{{\bf s}}^{\rm (pk)}(\theta_n)}
=\frac{\unity^{\otimes N}}{2^N},
\label{mixN}
\ee
\end{subequations}
while we obtain for the corresponding von Neumann  
entropy $S(e)=\sum_{j=1}^NS(\rho_{\rm pk}^{(j)})=N$.

The secrecy of the private key $d$ is guaranteed by the Holevo's theorem. 
In particular, let us denote by $I(x,d)$ the 
mutual information between the private key, and a variable containing the 
information an adversary (Eve) may have obtained by performing quantum 
measurements on the public key. Since the public-key qubits are prepared 
at random and independently in pure states, we have from Holevo's theorem  
$I(x,d)\leq S(e)=N$. Hence, $I(x,d)\ll H(d)$ provided 
\begin{subequations}
\label{conds}
\be
\log_2(|\tilde{\PosInt}|)+N\bar{n}\gg N,
\label{cond}
\ee
where $\bar{n}=(n_{\rm u}+n_{\rm l})/2$. Clearly, to satisfy  
condition (\ref{cond}) it is sufficient to have either 
$\bar{n}\gg 1$ or $\log_2(|\tilde{\PosInt}|)>>N$. In the protocol of the previous 
section, both of these requirements are fulfilled simultaneously 
since $n$ is chosen at random from the set of positive integers $\PosInt$ 
with the constraint $n>>1$. 
Hence, the inequality $I(x,d)\ll H(d)$ also holds that is, Eve's 
information gain is much smaller than the entropy of the private key $d$, 
which thus remains practically unknown to her. 
Accordingly, the conditional entropy $H(d|x)$ is given by 
$H(d|x)\equiv H(d)-I(x,d)\approx  H(d)$,  
which establishes the uniformity of the private key over 
$\Dr=\tilde{\PosInt}\times\Int_{2^n}$, after the measurements 
on the public-key state.

So, we have seen that by making the public key available to every one,  
we do not compromise the security of the protocol for $n\gg 1$, i.e., 
the public key may reveal only negligible information about the private key. 
When multiple copies of the public key, say $k$, are simultaneously 
in circulation, Eve's mutual information with the key increases, but is again 
upper bounded as follows $I(x,d)\leq Nk$. In this case, secrecy of 
the private key is always guaranteed if  
\be
\log_2(|\tilde{\PosInt}|)+N\bar{n}\gg Nk, 
\label{cond2}
\ee
\end{subequations}
which defines an upper bound on the number of copies of the public key that 
can be issued. This is in contrast to conventional public-key cryptosystems, 
where there are no such limitations. 

To summarize, the secrecy of the private key is guaranteed 
by the fact that the public key is quantum and unknown to every one  
except Alice. Moreover, the state of each public-key qubit is 
chosen at random and independently of the other qubit states.  
In other words, there is no redundancy or pattern in the public key, 
that could be explored by a potential adversary. 
Information gain on the state of the public key (and thus the private key),  
can be obtained only by performing measurements on the public-key qubits, 
at the expense of disturbing irreversibly their state. 
In any case, according to Holevo's theorem, this information gain cannot 
exceed one bit per qubit and thus, for $k$ copies of the public key 
simultaneously in circulation, the private key is secret as long as 
condition (\ref{cond2}) is satisfied. 
Furthermore, by virtue of the no-cloning theorem \cite{book3}, Eve cannot create 
additional copies of Alice's quantum public key, besides the copies 
provided by Alice or the KDC.  In particular, the fidelity of the clone for 
each public-key qubit is smaller than one \cite{ScaRMP05} and thus, the fidelity of the 
public-key clone drops exponentially with the key length $N$.

Finally, it is worth noting that according to the key-generation stage 
of Sec. \ref{secIII},  
there is a one-to-one correspondence between the private key and the 
public key. As a result, any information an adversary may obtain about 
the state of the $j$th public-key qubit $\ket{{\psi_{s_j}(\theta_n)}}_{j}$, 
is immediately associated with the the $j$th element of the private 
string ${\bf s}$. One may alter this situation, by applying a random 
permutation $\Pi$ on the public-key qubits, before they become 
publicly available.  
In this case, the $j$th element of the private string ${\bf s}$  
is mapped to the state of the 
$\Pi(j)$th qubit (i.e., $s_j\mapsto \ket{{\psi_{s_j}(\theta_n)}}_{\Pi(j)}$),
which is unknown to Eve if $\Pi$ is unknown. Hence, even if 
Eve were able to know precisely the state of each public-key qubit, 
she would have to guess the right permutation in order to deduce the 
private string ${\bf s}$. From another point of view, permuting the public-key 
qubits for a given private key is equivalent to  
preparing the public-key qubits in states determined by a permutation 
of the private string $\Pi({\bf s})$, which is unknown to Eve. 
In this case, the private key consists 
of three parts, i.e., $d^\prime=(n,{\bf s}, \Pi)$. The corresponding joint 
entropy is given by $H(d^\prime)=H(d)+H(\Pi|{\bf s},n)$, with $H(d)$ 
defined earlier. Accordingly, 
the left-hand side of Eqs. (\ref{conds}) increases by $H(\Pi|{\bf s},n)$, 
whereas the maximum information gain for a potential adversary 
is determined by the Holevo's bound and remains constant.   

In the following we analyze the security of our encryption scheme, against 
various types of attacks aiming at the recover of the plaintext and/or 
the private key, from the quantum ciphertext. These attacks are 
generalizations of the corresponding attacks on conventional asymmetric 
encryption schemes \cite{book2}.  In contrast, however, to their classical counterparts, 
in the quantum attacks Eve does not know the state of the quantum public 
key, but is allowed to perform arbitrary operations and measurements on it. 
The only assumption in the following analysis is that Alice's decryption 
device is manufactured so that is automatically deactivated when it 
performs $k$ consecutive decryptions on $N$-qubit states. In this way 
we guarantee that no more than $k$ copies of Alice's public key 
will be used. When these copies are exhausted, Alice must generate a new pair of 
keys $(e^\prime,d^\prime)$, and update accordingly her decryption device. 
To this end, the old private key may act as a quantum password, which ensures authorized 
access to the decryption device.

\subsection{Chosen-plaintext attack}
Typically, in a chosen-plaintext attack, Eve is allowed to obtain a number 
of plaintext-ciphertext pairs of her choice. More precisely, given $k$ copies of 
Alice's public-key state $\rho_{\rm pk}$, and $k$ plaintexts in binary form  
$\{{\bf a}_1, {\bf  a}_2, \ldots,{\bf a}_k\}$, with ${\bf a}_j\in\{0,1\}^{r_j}$ 
and $r_j\leq N$, she obtains a sequence of cipher states  
$\{\rho_{\rm e}^{(1)}, \rho_{\rm e}^{(2)}, \ldots, \rho_{\rm e}^{(k)}\}$,
where 
\[\rho_{\rm e}^{(j)}=\hat{\tilde{{\cal R}}}_{{\bf a}_j}^{(r_j)}(\pi) \rho_{\rm pk} 
\hat{\tilde{{\cal R}}}_{{\bf a}_j}^{(r_j)\dag}(\pi).
\]
The collective rotation on $r_j$ qubits is defined as  
\be
\hat{\tilde{{\cal R}}}_{\bf x}^{(r_j)}(\varphi)\equiv 
\bigotimes_{i=1}^{r_j} \hat{\cal R}^{(i)}(x_i \varphi). 
\label{colR}
\ee
Subsequently, Eve may explore her database, in order to decrypt an unknown message 
encrypted with Alice's public key, or gain further information on Alice's private key. 
For the sake of simplicity, and without loss of generality, in the following we 
assume that $r_j=N,~\forall~ j$. 

Let us discuss first whether Eve can gain significant information, by 
encrypting plaintexts (i.e., obtaining cipher states) of her choice.  
As discussed in the previous subsection, Eve can obtain only negligible 
information about the private key, by performing measurements 
on the public-key qubits. Thus, for Eve the private key is unknown, and 
uniformly distributed over $\Dr$. Accordingly, the state of the 
public key $\rho_{\rm pk}$ is chosen at random from the ensemble 
$\{p(d),\ket{\Psi_{\bf s}^{(\rm pk)}(\theta_n)}\}$, and is thus given by Eq. (\ref{mixN}). 
Note now that this maximally mixed state remains invariant under Eve's rotations 
\footnote{Actually, the state (\ref{mixN}) remains invariant even under a general 
quantum operation (completely positive map) 
$\Er:\rho\to\Er(\rho)=\sum_i \hat{E}_i\rho \hat{E}_i^\dag$, 
with $\sum_i \hat{E}_i\hat{E}_i^\dag=\unity$.}, 
and thus any plaintext ${\bf a}_j$ is mapped to the same 
cipher state, i.e., ${\bf a}_j\mapsto\rho_{\rm pk}$.  
Hence, on average, there is no information gain for Eve. 
The same conclusion can be drawn on the basis of Holevo's theorem. 
In particular, since the state of the public key is unknown to Eve, the cipher state 
is also unknown to her. Hence, Eve can extract at most $Nk$ bits of information 
from measurements on all of the $k$ cipher states, which is negligible in view of condition 
(\ref{cond2}).

The remaining question is whether Eve can use her plaintext-ciphertext 
database, in order to decrypt Bob's message, which has been encrypted 
with the same public key. First of all, recall that Bob encrypts his message 
${\bf m}\in\{0,1\}^r$, by transforming the state of the public key 
as follows
\be
\rho_{\rm pk}\stackrel{{\bf m}}{\longrightarrow} 
\rho_{\rm c}=\hat{\tilde{{\cal R}}}_{{\bf m}}^{(r)}(\pi)  
\rho_{\rm pk}\hat{\tilde{{\cal R}}}_{{\bf m}}^{(r)\dag}(\pi).
\ee
As mentioned above, the mixed 
state $\rho_{\rm pk}$ remains invariant under these rotations, and thus 
all of the possible messages yield the same cipher state, 
i.e.,  $\rho_{\rm c}=\rho_{\rm pk}$. 
Hence, Eve cannot distinguish between distinct messages, and the 
encryption scheme under consideration is provably secure \cite{Boy03}.

Finally, note that a protocol which is secure against chosen-plaintext attacks, 
is also secure against less powerful attacks, such as the 
ciphertext-only and the known-plaintext attacks \cite{book2}.
In the following, we analyze the forward-search attack, that is 
a chosen-plaintext attack adapted to small message spaces.

\subsection{Forward-search attack}
The forward-search attack can be very efficient (at least 
for conventional cryptosystems) when the number of all possible messages is small.
In this case, Eve may obtain multiple copies of Alice's public-key, and create 
the ciphertexts corresponding to each possible message. 
Subsequently, she may try to deduce the encrypted message, by 
comparing the unknown ciphertext with the ciphertexts in her database. 

For the encryption scheme under consideration, however, the crucial information 
is not the actual angle of the rotation, but rather whether a public-key qubit 
has been rotated or not (see stage 2 in Sec. \ref{secIIIa}). 
Hence, instead of creating her own plaintext-ciphertext database, 
it is sufficient for Eve to compare the cipher state sent from Bob to Alice, 
with a copy of Alice's public-key. 

To analyze this attack, let us focus on an 1-bit message $m\in\{0,1\}$. 
Bob encodes his message by applying the rotation ${\cal R}(m\pi)$ on 
Alice's public-key qubit, which is prepared in a state $\ket{\psi_s({\theta_n})}$ 
chosen at random from $\Rat_n$, for some $n\gg 1$. 
To deduce Bob's message, Eve performs a SWAP test \cite{Buh01} between the cipher 
qubit sent from Bob to Alice, and a copy of Alice's public-key 
qubit. 
In this way, she will learn whether the cipher-qubit state has 
been rotated with respect to the state of the public-key qubit. 
Such a test, succeeds with average probability $p_{\rm suc}=3/4$. 
Moreover, at the end of the test the two qubits are entangled, and Eve cannot distinguish 
between them. Hence, she cannot compare Bob's cipher state  
with the public-key state  more than once.

Alice and Bob can reduce considerably $p_{\rm suc}$, by encoding the message on the state 
of two, or more public-key qubits. For instance, using two public-key qubits in the state 
$\ket{\psi_{s_1}({\theta_n})}_1\otimes\ket{\psi_{s_2}({\theta_n})}_2$, the message 
``0'' is encoded  by applying an operation randomly chosen 
from the set $\{\hat{\cal R}^{(1)}(0)\hat{\cal R}^{(2)}(0), 
\hat{\cal R}^{(1)}(\pi)\hat{\cal R}^{(2)}(\pi)\}$, 
whereas ``1'' is encoded using an operation from the set 
$\{\hat{\cal R}^{(1)}(0)\hat{\cal R}^{(2)}(\pi), 
\hat{\cal R}^{(1)}(\pi)\hat{\cal R}^{(2)}(0)\}$.   
Thus, to deduce Bob's message, Eve has to identify correctly the operations  
performed on both qubits. In this case, Eve succeeds with probability 
$p_{\rm suc}=(3/4)^2\approx 0.56$; that is, slightly better than random guessing.  
In general, when each bit of a message is encoded to $\alpha$ qubits, 
Eve has to perform $\alpha$ successive SWAP tests to deduce it,  
and the average success probability is $(3/4)^\alpha$; that is 
worse than random guessing for $\alpha>3$. 

In the forward-search attack discussed above, Eve performs independent (individual) SWAP 
tests between the corresponding qubits of the cipher state and a copy of the public key. 
The question arises here is whether Eve may increase her probability of success, 
by performing collective measurements on all the qubits of the cipher state 
and the public key. This issue deserves further investigation, 
and will be addressed elsewhere. Nevertheless, the mere fact that each  
public-key qubit is prepared at random and independently of the others, 
suggests that the optimal attack (i.e., the attack that maximizes 
Eve's probability of success), involves only individual measurements on 
various qubit pairs, consisting of the corresponding qubits of the cipher 
state and the public key. 
In particular, as discussed in Sec. \ref{secIVB}, there is no redundancy or pattern 
in the public key (and thus in the cipher state) which could be explored 
in a collective measurement.  

\subsection{Chosen-ciphertext attack}
In this scenario, Eve has access to Alice's decryption device, but not to the private 
key. Providing judiciously chosen cipher states, she receives the corresponding plaintexts. 
The only restriction is that Alice's device does not allow more than $k$ decryptions 
on $N$-qubit states with the same private key.
As before, Eve's objective is to deduce the private key, or decrypt Bob's 
message at a later instant, when she does not have access to the decryption device. 

The chosen-ciphertext attack can be analyzed along the lines of the 
previous sections. Let us discuss briefly, for instance, 
the security of the private key. 
In a chosen-ciphertext attack Eve can prepare arbitrary 
multi-qubit states, not necessarily related to the public key.
For instance, Eve may ask for the decryption of an 
$N$-qubit state $\rho_{\rm e}$,  
where the qubits are entangled among themselves as well as with 
another ancillary system. Nevertheless, as soon as the qubits are input to 
the decryption device, Eve has no access to them. 
First, the decryption device undoes the initial rotations  
on the qubits, as determined by the private key $d$. 
For Eve, who does not have access to the private key, the input state 
is transformed to a state $\rho_{\rm e}^\prime$ randomly chosen from the ensemble 
$\{p(d), \hat{\tilde{{\cal R}}}_{\bf s}^{(r)\dag}(\theta_n)
\rho_{\rm e} \hat{\tilde{{\cal R}}}_{\bf s}^{(r)}(\theta_n)\}$, i.e., 
\be
\rho_{\rm e}\stackrel{d}{\longrightarrow}  \rho_{\rm e}^\prime
=\sum_{d\in\Dr} p(d) \hat{\tilde{{\cal R}}}_{\bf s}^{(r)\dag}(\theta_n)
\rho_{\rm e} \hat{\tilde{{\cal R}}}_{\bf s}^{(r)}(\theta_n),
\ee
with the collective rotations given by Eq. (\ref{colR}). 
Eve learns only the outcomes of the projective measurements 
performed at the end of the decryption stage. According to 
Holevo's theorem, however, these outcomes cannot provide her with more than 
$N$ bits of classical information about the private key. 
Of course Eve has the chance to perform up to $k$ such 
decryptions, but as long as condition (\ref{cond2}) is satisfied,  
her information gain is not sufficient to determine the private key.

\section{Discussion} 
\label{secC}
In conclusion, we have discussed cryptographic applications of single-qubit 
rotations in the framework of quantum trapdoor (one-way) functions. 
We also demonstrated how such a function can be used as a basis for 
a quantum public-key cryptosystem, whose security, 
in contrast to its classical counterparts, 
relies on fundamental principles of quantum mechanics.  
More precisely, in the proposed encryption scheme, 
each user creates a key consisting of two parts: 
a private key, which is purely classical, and a public key, which involves a 
number of qubits prepared independently in states specified by the 
private key. The sender encrypts his message on the recipients public key 
by rotating the state of its qubits. A potential adversary cannot deduce the 
encrypted message without knowing the recipient's private key.  

One might have noticed here external similarities of the proposed encryption 
scheme to the Y00 protocol \cite{Bar03}.
To avoid any misunderstandings, we would like to point out some crucial differences 
between the two schemes. 
First of all, the security of the Y00 protocol is claimed to rely on quantum noise 
which renders the discrimination of closely spaced mesoscopic states impossible. 
On the contrary, the security of the proposed public-key encryption scheme relies on 
the Holevo's bound and the no-cloning theorem. 
Second, the Y00 is a symmetric encryption scheme whereas the present work 
involves asymmetric cryptosystems (different keys are used for encryption and 
decryption). Third, in the Y00 protocol the two legitimate users share a short 
secret key in advance, which is expanded in the course of the protocol. No 
secret information is necessary for the functionality of the present protocol. 

Various security issues pertaining to the proposed asymmetric encryption scheme, 
have been analyzed in the context of a futuristic scenario, where all of 
the entities participating in the cryptosystem possess quantum computers, 
and are connected via ideal quantum channels. There are various questions 
yet to be explored, especially in connection with the extension of the present 
ideas to more realistic scenarios, where the legitimate users are limited by 
current technology. For instance, in the presence of a lossy quantum channel, 
quantum error-correction codes can be used to increase the robustness of the protocol.  
We have already seen that by encoding 1 bit on two qubits we make the encryption more 
robust against the forward-search attack.  

In any case, the purpose of the present work was to introduce certain basic ideas 
underlying quantum public-key encryption, and set an appropriate theoretical framework. 
We also demonstrated how fundamental properties of quantum systems 
and certain theorems of quantum mechanics may provide a barrier, due to  
complexity of effort, between legitimate users and adversaries, which 
is the cornerstone of quantum public-key encryption. 
We hope that our results and discussion will stimulate further investigations 
on these topics, so that light is shed on crucial questions,  
pertaining to the power and the limitations of asymmetric quantum cryptography. 
Moreover, such investigations might lead to the development of practical 
public-key encryption schemes, or other provably secure quantum cryptographic 
primitives (e.g.,  digital signatures, hash functions, etc).

\section{Acknowledgments} 
I am grateful to S. J. van Enk and P. Lambropoulos for helpful comments and discussions. 
The work was supported in part by the EC RTN EMALI (contract No. MRTN-CT-2006-035369).

\end{document}